\documentclass[twocolumn]{cinc}
\usepackage{graphicx}
\begin{document}
\bibliographystyle{cinc}

\title{Assessing Measures of Atrial Fibrillation Clustering via \\Stochastic Models of Episode Recurrence and Disease Progression}

\author {Julie Eatock$^1$, Yen Ting Lin$^2$, Eugene TY Chang$^3$, Tobias Galla$^2$, Richard H Clayton$^3$\\
\ \\
{\normalsize$^1$ Department of Computer Science, Brunel University London, Uxbridge, UK\\
$^2$ Theoretical Physics, School of Physics and Astronomy, The University of Manchester, Manchester, UK\\
$^3$Insigneo Institute for in-silico Medicine and Department of Computer Science, \\University of Sheffield, Sheffield, UK
}}

\maketitle

\begin{abstract}
Atrial fibrillation (AF) is a leading cause of morbidity and mortality. AF prevalence increases with age, which is attributed to pathophysiological changes that aid AF initiation and perpetuation. Current state-of-the-art models are only capable of simulating short periods of atrial activity at high spatial resolution, whilst the majority of clinical recordings are based on infrequent temporal datasets of limited spatial resolution. Being able to estimate disease progression informed by both modelling and clinical data would be of significant interest.  In addition an analysis of the temporal distribution of recorded fibrillation episodes \emph {AF density} can provide insights into recurrence patterns. We present an initial analysis of the AF density measure using a simplified idealised stochastic model of a binary time series representing AF episodes. The future aim of this work is to develop robust clinical measures of progression which will be tested on models that generate long-term synthetic data. These measures would then be of clinical interest in deciding treatment strategies.

\end{abstract}
%
\section{Introduction}
Atrial Fibrillation (AF) is a progressive condition characterised by an irregular dissynchronous atrial contraction. It is non-symptomatic and intermittent and it is often difficult to differentiate between the various states of AF \cite{L-11}. Initially, AF may spontaneously terminate after a short duration, but as the disease progresses, each episode may increase in length and inter-episode times may gradually decrease, eventually leading to long episodes lasting weeks or more \cite{K-05}.

AF prevalence increases with age, which is attributed to pathophysiological changes associated with AF initiation and perpetuation \cite{Co-14}. Recording and modelling atrial electrical activity are used to identify and inform hypotheses surrounding AF mechanisms, although current biophysical models, which are computationally demanding, can only simulate short periods of atrial activity, whilst the majority of clinical recordings are based on infrequently acquired datasets of the 12 lead ECG is of limited spatial resolution. Treatment and diagnosis of AF are based on the duration of the AF events themselves, the likelihood of spontaneous conversion to sinus rhythm, and the likelihood of a successful intervention. 

To assess the likelihood of these outcomes clinicians have to assess and classify the progression of AF. There are a range of measures in common use, including statistics of mean and median AF durations, number of episodes and AF burden, the time in AF over the observation window. In general, AF episodes are hard to diagnose and predict due to their intermittent nature and because temporal persistence of episodes do not generally follow a distribution. The ability to predict the increased susceptibility of a patient to future AF episodes and estimate disease progression informed by both modelling and clinical data would be of significant interest.  

Recent studies by Charitos et al \cite{C-13,C-14} sought to classify AF patients based on temporal analysis of their recorded fibrillation episodes and introduced a novel measure of \emph{AF density} to characterise AF recurrence patterns. The premise of the measure is to better capture temporal AF persistence than existing measures. AF density evaluates the ratio of the cumulative deviation of the patient's actual AF burden (time in AF over monitored time) from the hypothetical uniform burden development, to that of a maximal possible burden aggregation from the hypothetical uniform burden development. We present an initial analysis of the AF density measure using a simplified idealised stochastic model of a binary time series representing AF episodes, with the waiting times between two successive `event's drawn from a Gamma distribution. 

\section{Methods}
The AF density is defined by Charitos et al \cite{C-13} as follows:\\
\emph{For a patient with a total AF burden $b$ (expressed as the proportion of the observation time the patient is in AF), who is monitored for time $T$, we denote the minimum contiguous monitored time throughout the monitored period $T$ required for the development of a proportion $p$ of the patient's total observed burden ($b$) as $T(p;b)$. This time, expressed as proportion of the total observed time $T$, is $F(p;b) = T(p;b)/T$. The cumulative deviation of the patients actual burden development from the hypothetical uniform burden development can be evaluated as}

\begin{equation}
\int^1_0 |F(p;b)-p |\,dp.
\end{equation}
\emph{For the hypothetical patient with maximum temporal aggregation of burden $b$ (the complete burden as one continuous AF episode) the cumulative deviation of this patient's burden development from the hypothetical uniform burden development is evaluated as $(1-b)/2$. AF density is then defined as:}
\begin{equation}
2\frac{\int^1_0 |F(p;b)-p |\,dp}{1-b}.
\end{equation}

A sensible statistical measurement must be well-defined and must have a clear operational definition with clear corresponding clinical relevance. In the definition of Charitos et al, the overall time of observation is always normalised to be 1, which, mathematically, can result in ill-defined measure. When the duration of AF episodes corresponding to inter-episode times have a natural time scale, this measure may not be sensible without specifying the length of the observation period (akin to giving a mean without also stating a standard deviation).\\

To illustrate this we generated binary time series data that was either in state 1 (on) or 0 (off). Instead of normalising by the total time $T$, we normalise the time such that total time of the patient in AF is 1 unit.\\

When in state 0, the time unit where an instantaneous transition to state 1 would occur was drawn from a Gamma distribution $Gamma(5,1/25)$. When in state 1, the time unit where an instantaneous transition to state 0 would occur was drawn from a Gamma distribution $Gamma(5,1/10)$. This was repeated to obtain a time series of 350 time units representing days. The theoretically predicted value for the burden defined as the time in state 1/overall time was  $~T_{on}/(T_{on}+T_{off})=\lambda_{off}^{-1}(1/\lambda_{off}+1/\lambda_{on})^{-1}\approx 0.2857$. 

\section{Results and Discussion}
\begin{figure}[h!]
\begin{center}
	\includegraphics[width=0.5\textwidth]{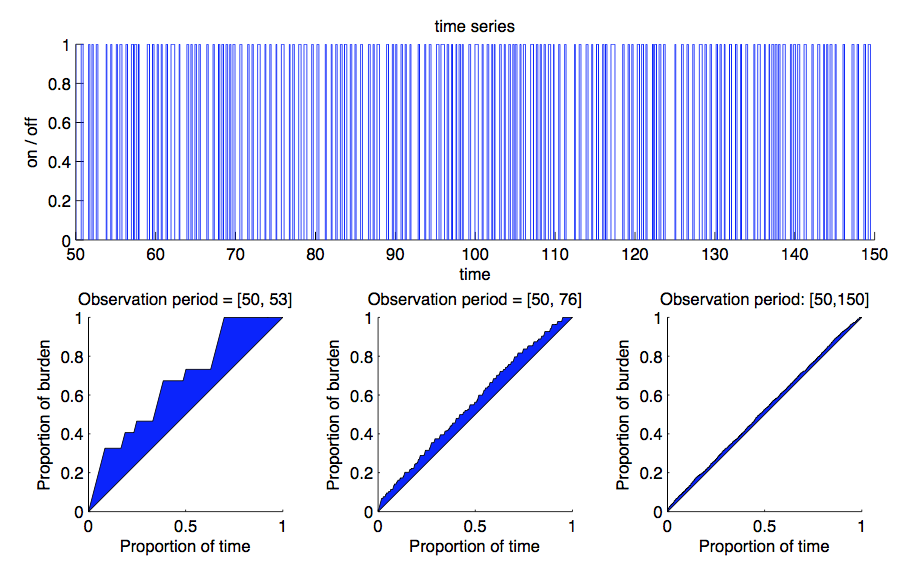}
	\caption{Visualisation of simulated AF episodes using transition rates drawn from a Gamma distribution. The burden remains relatively static with no significant change in the time series of the same period (top), but the AF density (bottom panel) decreases from as the observation window is altered from 3 days to 100 days\label{fig:AF1}.}
\end{center}
\end{figure}

Since AF density requires a specific period of observation, we measured the burden and density for the time series using three different time windows all starting at the same point [50,53], [50,76], [50,150].  From Figure \ref{fig:AF1} we can see that while the burden remains static (upper panel) despite the length of the observation period, the density decreases as the observation window is increased from 3 days to 100 days. This is illustrated graphically as the shaded area bound between the cumulative proportion of burden and the diagonal line representing uniform burden (Figure \ref{fig:AF1} lower panels) - despite there being no significant change in the time series over the same period  

\begin{figure}[h!]
\begin{center}
	\includegraphics[width=0.5\textwidth]{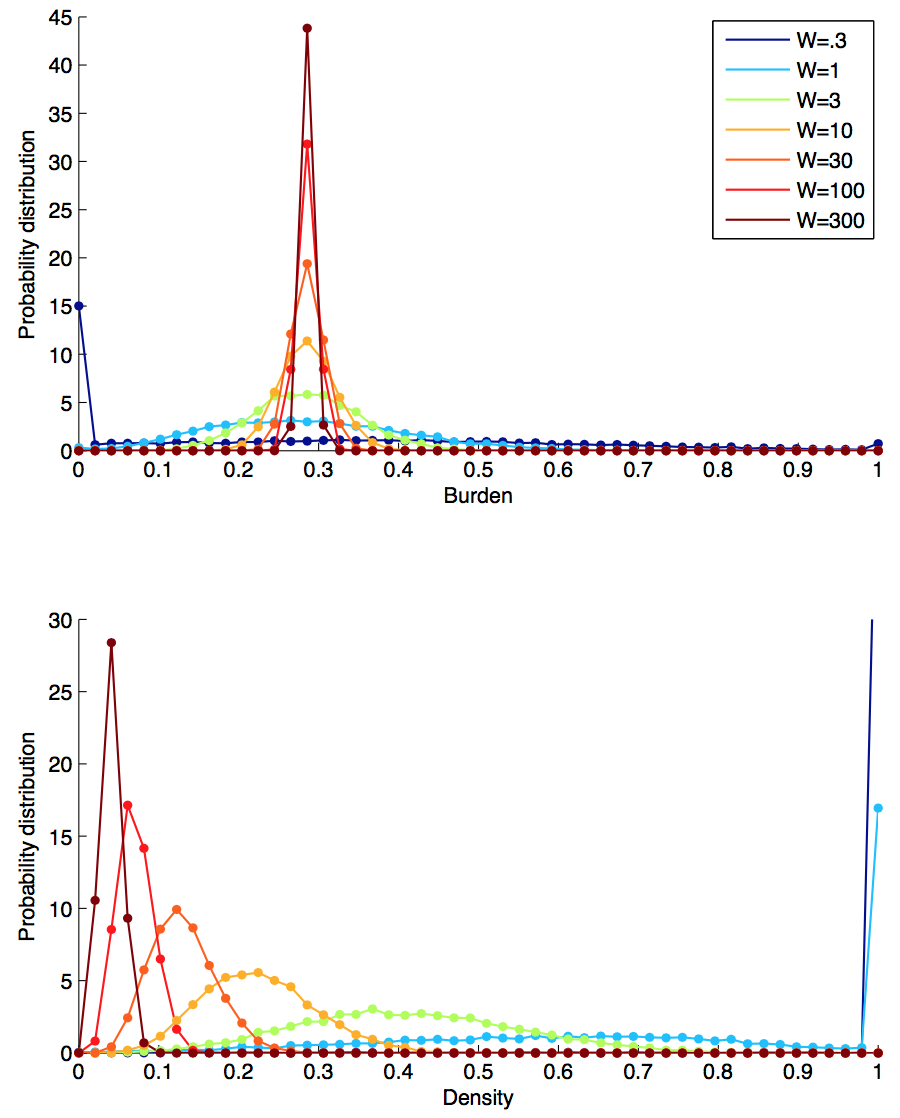}
	\caption{Burden (top) and density (bottom)histograms for different observation periods W =[0.3; 1; 3; 10; 30; 100; 300]. The burden showns a sharper peak around the theoretical burden as W increases whilst the density tends to zero.\label{fig:AF2}.}
\end{center}
\end{figure}

Examining this further we randomly generated $10^4$ sets of time series data and measured the burden and density for the time series in between [50; 50 + W] where the W is length of the observation period W =[0.3; 1; 3; 10; 30; 100; 300]. Figure \ref{fig:AF2} shows that while the burden shows a sharper peak around the theoretical burden as the observation period increases, the density tends to zero as the observation period increases, highlighting its sensitivity to the observation period.

\begin{figure}[h!]
\begin{center}
	\includegraphics[width=0.5\textwidth]{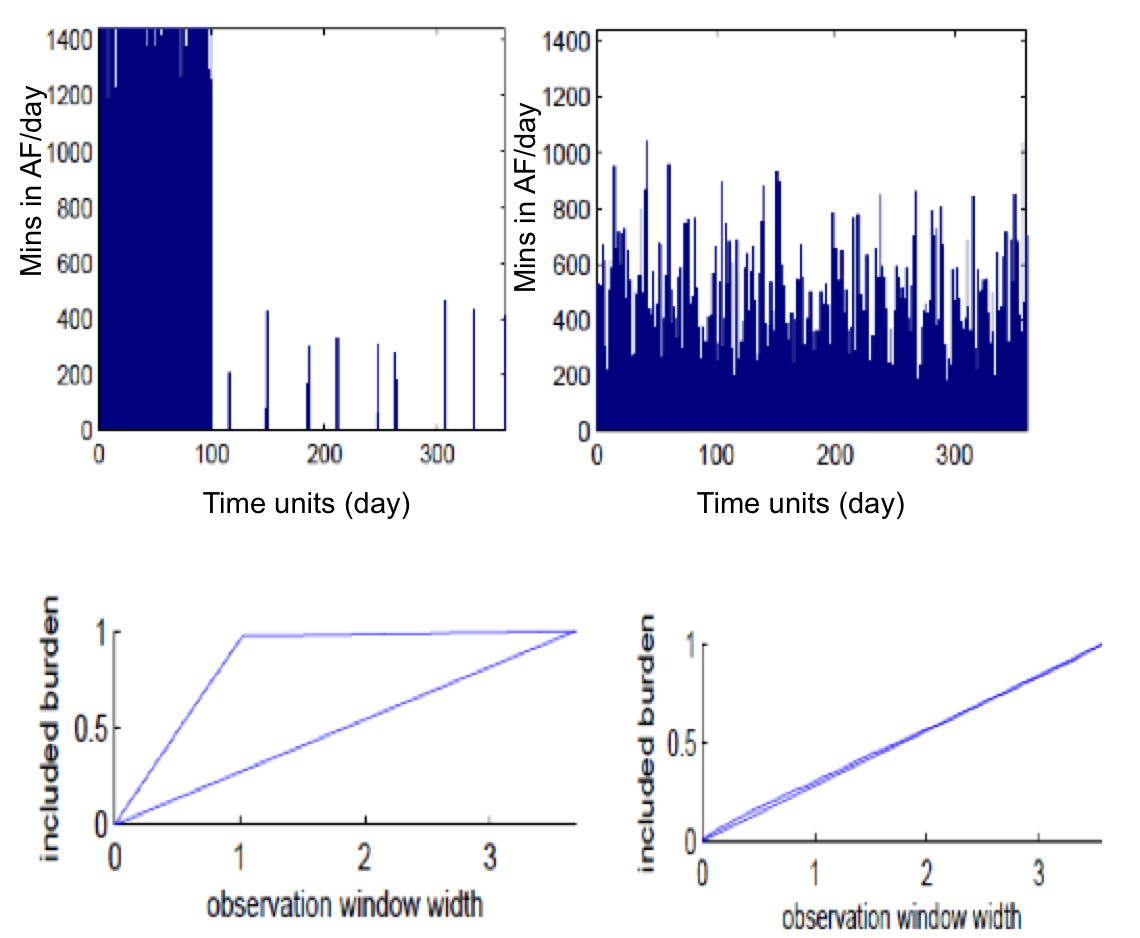}
	\caption{AF density plotted for two patients of similar burden (0.27115 and 0.28102) but with different profiles. Patient A has most of the episodes occurring early (top left) which leads to a high AF density of $0.9538$ (bottom left) which significantly deviates from uniform burden, whilst patient B experiences a more even distribution of AF over the recording period (top right), with a density of $0.033297$ differing little from uniform burden (bottom right)\label{fig:AF5}.}
\end{center}
\end{figure}

To illustrate the effects of this when applied to AF in a clinical setting, we generated data that displayed similar patterns to the examples illustrated in Charitos' paper, and again calculated the burden and density associated with the time series. Patient A experiences nearly their entire AF burden (i.e. time spent in AF) in 80 days, while patient B experiences their AF more evenly distributed across the entire recording period.  Fig \ref{fig:AF5} show the two time series over the period of 1 year with similar AF burden ((a) = 0.27115, (b) = 0.28102), but markedly different AF density profiles ((a)=0.9538 (b)=0.033297). This illustrates why AF burden is insufficient as a stand alone measure and that Charitos et al's density measure  is very effective at capturing the difference in the patient's AF profile over a stated time period. \\

\begin{figure}[h!]
\begin{center}
	\includegraphics[width=0.5\textwidth]{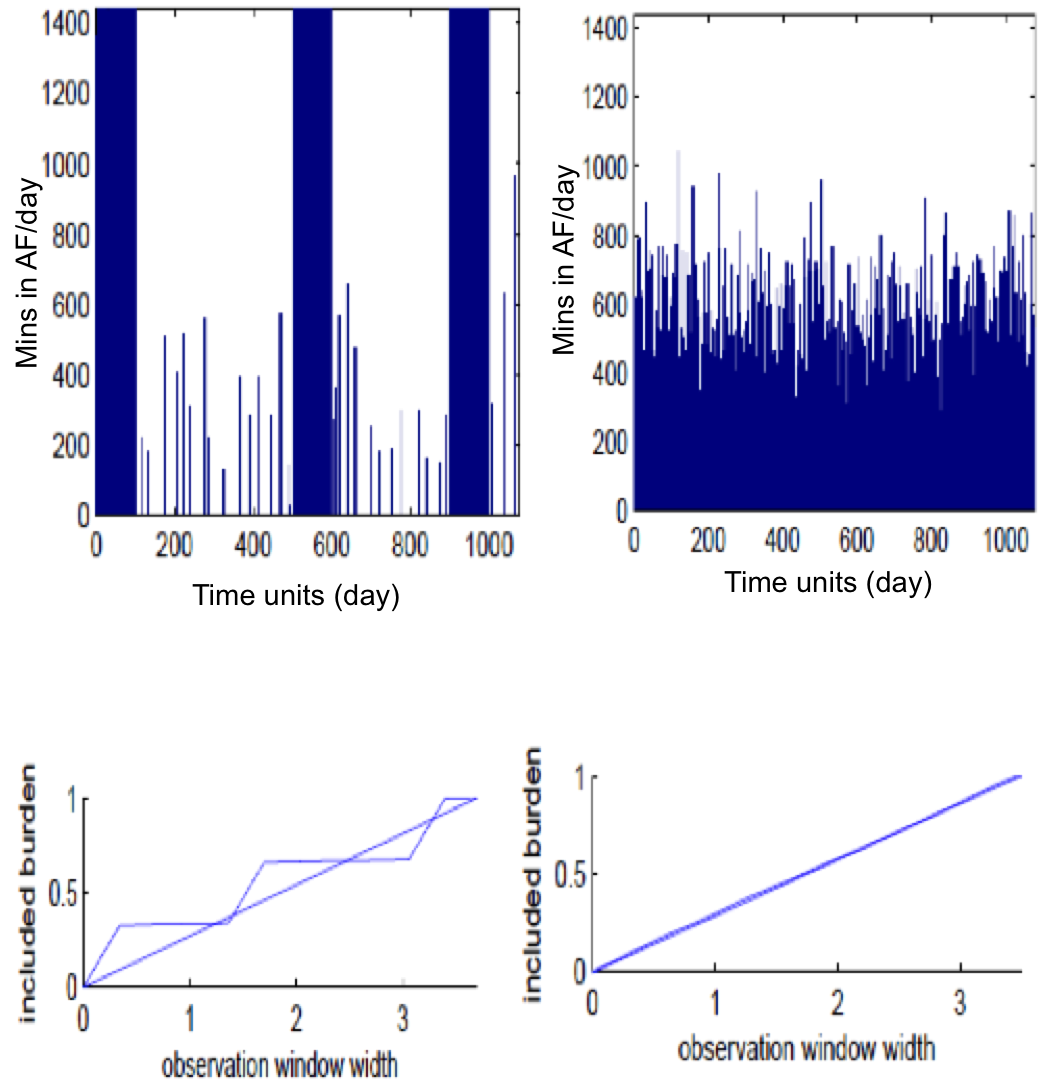}
\caption{We expand the time series to three years, such that patient A (left) has 3 episodes of AF with long periods in between with short AF events whilst patient B (right) has as a spread of AF events. Using the AF density measure, patient A now has AF density of 0.24763 (bottom left) as the quiescence periods between the long events reduce the deviation from uniform burden. Patient B has density 0.025055 (bottom right) which is very close to uniform burden. Both patients still have AF burden of 0.27073 (A) and 0.28555 (B) \label{fig:AF6}.}
\end{center}
\end{figure}

If we now expand the same time series to three years (Figure \ref{fig:AF6}) patient A has 3 distinct episodes of continual AF, with very long periods between with a very low number of AF events. Patient B has AF events every day that last between about 5 and 15 hours. If again we obtain the burden and density, as expected we find the burden between 1-year and 3-year time series and between the two patients themselves, remains similar (a)=0.27073, (b)=0.28555.
The density for patient B over the 3-year time series (0.025055) is similar to the density over the 1-year period (0.033297) but still reduced, despite the very similar profile. Patient A had a density of 0.9538 at the 1-year time period, while the 3-year observation period reveals a much lower density of 0.24763. Using this measure in a clinical setting could misleadingly imply that something in the pattern of AF episodes has changed significantly, to where the AF events were more uniformly distributed, whereas in reality this would not be the case.\\

This raises the question of the requirements of a robust measure of temporal nature of AF progression, particularly as we know that events can be highly clustered and asymptomatic and depending on the recording strategy can easily missed.

\section{Conclusion, Limitations and Further Work}
We have demonstrated using a simplistic synthetic time series that we need to consider the quality and robustness of the clinical measures AF burden and AF density. We conclude that, whilst a measure of temporal dispersion is necessary to describe differences in AF persistence, and density as defined by Charitos et al captures this well in a fixed and stated time period, it is sensitive to the size of the observation window which is a limitation that needs to be addressed before use in a clinical setting. The criteria that these measures need to meet in order to accurately assess progression over a moving or expanding time window still needs to be fully identified.\\
We believe that any clinical measures need to be fully tested on high quality synthetic data where the distribution of AF episodes is fully known, before being applied to real clinical recordings. A significant limitation of this study is that the generated time series used for the present analyses was not physiological and did not exhibit any of the characteristic progressions more commonly associated with AF. This motivates the need for more physiologically-representative models of AF, which can simulate patterns of AF episodes and subsequent progression over time. With such models, we can then develop and test robust measures of progression of AF which could be applied in clinical diagnosis and treatment.

\section*{Acknowledgements}  
We acknowledge funding by the Engineering and Physical Sciences Research Council (UK), grant EP/K037145/1. ETYC acknowledges a PDRA award from the Network for Computational Statistics \& Machine Learning, UK
\balance

\begin{correspondence}
Julie Eatock\\
Dept. of Computer Science,\\
Brunel University London, \\
Uxbridge, UB8 3PH, \\
United Kingdom\\
Julie.Eatock@brunel.ac.uk

\end{correspondence}


\begin{thebibliography}{99}{ 

\bibitem{L-11}Lewis, M., Parker, D., Weston, C., Bowes, M. Screening for atrial fibrillation: sensitivity and specificity of a new methodology. Br. J. Gen. Pract. 61, 38-39. 2011.


\bibitem{K-05}Kerr, C.R., Humphries, K.H., Talajic, M., Klein, G.J., Connolly, S.J., Green, M., Boone, J., Sheldon, R., Dorian, P., Newman, D. Progression to chronic atrial fibrillation after the initial diagnosis of paroxysmal atrial fibrillation: Results from the Canadian Registry of Atrial Fibrillation. Am. Heart J. 149, 489-496. 2005. 


\bibitem{Co-14}Corradi, D. Atrial fibrillation from the pathologist's perspective. Cardiovascular Pathology 23, 71-84, 2014.


\bibitem{C-13}Charitos, EI, Ziegler, PD, Stierle, U, Sievers, HH, Paarmann, H, Hanke, T.  Atrial fibrillation density: A novel measure of atrial fibrillation temporal aggregation for the characterization of atrial fibrillation recurrence pattern. Applied Cardiopulmonary Pathophysiology 17: 3-10, 2013.


\bibitem{C-14}Charitos, EI, PŸrerfellner, H, Glotzer, TV, Ziegler, PD. Clinical classifications of atrial fibrillation poorly reflect its temporal persistence: Insights from 1,195 patients continuously monitored with implantable devices. J. Am. Coll. Cardiol. 63, 2840-2848,  2014.


}\end{thebibliography}
\end{document}